\documentclass {osa-article}

\journal{osac}
\newcommand{\blue}{\textcolor{blue}}

\articletype{Research Article}
\begin{document}

\title{Giant enhancement of THz wave emission under  double pulse excitation of thin water flow}

\author{Hsin-hui Huang,\authormark{1} Takeshi Nagashima,\authormark{2*} Tetsu Yonezawa,\authormark{3,4} Yasutaka Matsuo,\authormark{5} Soon Hock Ng,\authormark{6}  Saulius Juodkazis,\authormark{6,7,8*} and Koji Hatanaka\authormark{1,9,10*}}

\address{\authormark{1}Research Center for Applied Sciences, Academia Sinica, Taipei 11529, Taiwan\\
\authormark{2}Faculty of Science and Engineering, Setsunan University, \mbox{Osaka 572-8508, Japan}\\
\authormark{3}Division of Materials Science and Engineering, Faculty of Engineering, Hokkaido University, Hokkaido 001-0021, Japan\\
\authormark{4}Institute for the Promotion of Business-Regional Collaboration, Hokkaido University, Hokkaido 001-0021, Japan\\
\authormark{5}Research Institute for Electronic Science, Hokkaido University, Hokkaido 001-0021, Japan\\
\authormark{6}Nanotechnology Facility, Center for Micro-Photonics, Swinburne University of Technology, Melbourne, VIC~3122, Australia\\
\authormark{7}World Research Hub Initiative (WRHI), School of Materials and Chemical Technology, Tokyo
Institute of Technology, 2-12-1, Ookayama, Meguro-ku, Tokyo 152-8550, Japan\\
\authormark{8}Institute of Advanced Sciences, Yokohama National University, 79-5 Tokiwadai, Hodogaya-ku, Yokohama 240-8501, Japan\\
\authormark{9}College of Engineering, Chang Gung University, Taoyuan 33302, Taiwan\\
\authormark{10}Department of Materials Science and Engineering, National Dong-Hwa University, Hualien 97401, Taiwan}

\email{\authormark{*}t-nagash@mpg.setsunan.ac.jp (T.N.), \authormark{*}sjuodkazis@swin.edu.au (S.J.), \authormark{*}kojihtnk@gate.sinica.edu.tw (K.H.)}%% email address is required


\begin{abstract}
Simultaneous measurements of THz wave and hard X-ray emission from thin and flat water flow when irradiated by double femtosecond laser pulses (800 nm, 35~fs, transform-limited, 0.5 kHz, delay times up to 15~ns) were carried out at the transmission and the reflection sides of the flow with THz time-domain spectroscopy and Geiger counters. THz wave emission spectra show their dynamic peak-shifts toward the low frequency with the highest intensity enhancements more than $1.5\times 10^3$ times in |\textbf{\textit{E}}|$^2$ at the delay time of 4.7~ns between the two pulses, while X-ray intensity enhancements are limited to about 20 times at 0~ns under the same experimental conditions. The mechanisms for the spectral changes and the intensity enhancements in THz wave emission are discussed from the viewpoint of laser ablation on the water flow induced by the pre-pulse irradiation.
\end{abstract}

%%%%%%%%%%%%%%%%%%%%%%%%%%  body  %%%%%%%%%%%%%%%%%%%%%%%%%%
\section{Introduction}

THz science and technology~\cite{MittlemanJAP17} are continuously growing since the early report on THz wave emission~\cite{Fattinger89}. Applications have been  demonstrated in imaging~\cite{mittleman2018} with optical component developments~\cite{Sahoo2019}, spectroscopy for mesoscopic structures~\cite{Zeitler}, coherent measurements of intense THz wave~\cite{Mou19}, gas-detection in Fabry-P\'erot cavity~\cite{Hindle19}, tunable/compact laser~\cite{Chevalier19}, sensing for medical uses~\cite{Sun18}, and next generation 6G wireless communication~\cite{6G}. Recent progresses include studies under scanning tunneling microscopy~\cite{Hegmann13, Mittleman17, Hillenbrand17} with nanoscale spatial resolution. For further developments in the field, innovations in THz wave sources are expected.

One of the experimental methods for THz wave emission, apart from the well-developed techniques with semiconductors such as ZnTe and others \cite{Wilke07}, is through femtosecond laser-based plasma formation \cite{Hamster94} since much higher incident laser intensity can be put into the plasma formation. While solid samples are irreversibly damaged under such intense laser irradiation conditions, air~\cite{hochstrasser2000, Andreeva16} and noble gases (He, Ne, Ar, Kr, and Xe)~\cite{Zhang07, sakabe19} have been used as targets for plasma-based THz emission, thus circumventing the issue. Laser excitation with chirped pulses ~\cite{Savelev18} or two-colors~\cite{hochstrasser2000, Kress04, Zhang2006, Babushkin10, Savelev18} has been carried out with discussions mainly via the four-wave mixing or ponderomotive force mechanism. Furthermore, a feedback method of laser wave-front manipulation can control THz wave emission~\cite{Krushelnick17}. One disadvantage of gaseous targets is low absorption cross section at the incident laser wavelength. Considering the damage issue in solid samples, liquids can be an ideal target for plasma-based THz wave emission since it can be refreshed continuously during experiments. Indeed, experiments with a water film \cite{Tcypkin19, Tcypkin19Kerr, ZhangAPL17, Zhang18, ZhangAPL18} or a line \cite{Zhang19} and other liquids such as deionized-water, alcohols, acetone, dichloroethane, or carbon disulfide \cite{DeyNatCommun2014} have been carried out. It is also reported that THz wave emission can be induced in association with X-ray emission from thin water flow and its intensity is enhanced under double\blue{-}pulse excitation \cite{Hatanaka2018THz}.

%THz emission has been reported by Auston, Mourou, Fattinger and Grischkowsky since the 70s\cite{Auston75, Mourou81, Fattinger89, Grischkowsky90}. Since then it has been applied in medical sensing \cite{Sun18}, chemical analysis \cite{Zeitler}, or even more advanced applications such as THz scanning tunnelling microscope at a 0.3 nm spatial resolution \cite{Hegmann17, Hegmann13, Mittleman17, Hillenbrand17, Huber16}

%\textcolor{red}{laser-plasma-based THz emission}

%\textcolor{red}{Several ways of THz wave generation have been developed in recent years. Other than the conventional semiconductor crystals such as GaSe and ZnTe as the emitter, THz wave can also be generated from nonlinear propagation effects such as plasma interaction with different matters. As reported by Cook \textit{et al.} THz wave can be generated from laser pulses that is composed of a fundamental and second-harmonic (SH) lights and focused in air \cite{Hochstrasser2000}. In recent years, Krushelnick \textit{et al.} has made it possible for THz wave tunning and optimize THz signal in desired wavelength range.} 

%\textcolor{red}{Apart from gases as the target matter, there are groups around the world starting to use liquids as the target for THz wave generation. Zhang \textit{et al.} has been demonstrating THz wave generation from water film and water line, they have also reported that with the four-wave mixing technique THz energy is enhanced by two-orders of magnitude compared to just with single fundamental light \cite{ZhangAPL18}}

Water, among various liquids, is considered to be one of the best targets for intense laser irradiation for THz wave emission without serious degradation. Even at high laser repetition rates, liquid target can be utilised. Indeed, experiments with distilled water and other solutions were reported recently as described, discussions started on the emission mechanisms and characteristics of THz source. Similar to THz wave emission at meV-photon energies, X-ray emission at keV from water and aqueous solutions with solutes or additives such as electrolytes \cite{hatanaka08, hatanaka2002water} and gold nano-particles \cite{Masim2016, Hiro2018} has been also reported. Water as a universal solvent is a good target for investigation of mechanisms  and for development of THz wave sources for bright emission.

Double\blue{-}pulse excitation is another method for the THz wave emission enhancements \cite{nagashimaapl2017, Hatanaka2018THz, ZhangDP19, Ponomareva19}. Under double\blue{-}pulse excitation of Ar gas  at  delay times of 500 ps, THz emission increases 10-times as compared with single\blue{-}pulse excitation~\cite{nagashimaapl2017}. In the case of flat water flow\blue{,} when the thickness of jet was 100-240 $\mu$m  at the delay time of $\sim$6~ps, THz emission increases 4-times \cite{Ponomareva19}, while in the case of water jet of 210 $\mu$m diameter at the delay time of 50 ps, THz pulse energy increases more than 8-times~\cite{ZhangDP19}. For thinner flat water flow with 20 $\mu$m thickness, the enhancement was about 100-times in |\textbf{\textit{E}}|$^2$ estimated from TDS signal peak intensities at the nanosecond delays~\cite{Hatanaka2018THz}. As the related-mechanisms have been already discussed in X-ray emission from aqueous solutions \cite{hatanaka08, Hsu2017}, the pre-pulse irradiation plays an important role in pre-plasma formation and in dynamic morphological changes due to ablation within picosecond to nanosecond time delays. Water surface can be controlled on-demand by pre-pulse irradiation for plasma formation leading to solvated electron formation in water \cite{Bruggeman2016} and for transient surface structure formation by ablation~\cite{hatanaka08}. However, better understanding of the detailed mechanisms of THz wave emission and its enhancement are still indispensable especially in the field of strongly localised light-matter interactions occurring at focal volumes from nano-scale comparable with the wavelength of the incident laser pulses to micro-scale comparable with the wavelength of resultant THz wave emission.

In this paper, we report simultaneous measurements for THz wave and X-ray emission from thin water flow under double-pulse excitation with delay times up to 15~ns. Finely-synchronized THz wave and X-ray radiation is required for their complimentary usage, for instance, for imaging~\cite{ZhangNatPhoton2017}. This study explores the enhancement of THz wave radiation from the simultaneous THz/X-ray emission source which were first reported in gas, glass, and water~\cite{balakin2016, Hatanaka2018THz, Hamster1993}.

\section{Experimental Setup}
\subsection{Double pulse excitation of water flow}
\begin{figure}[tb]
\centering\includegraphics[width=11cm]{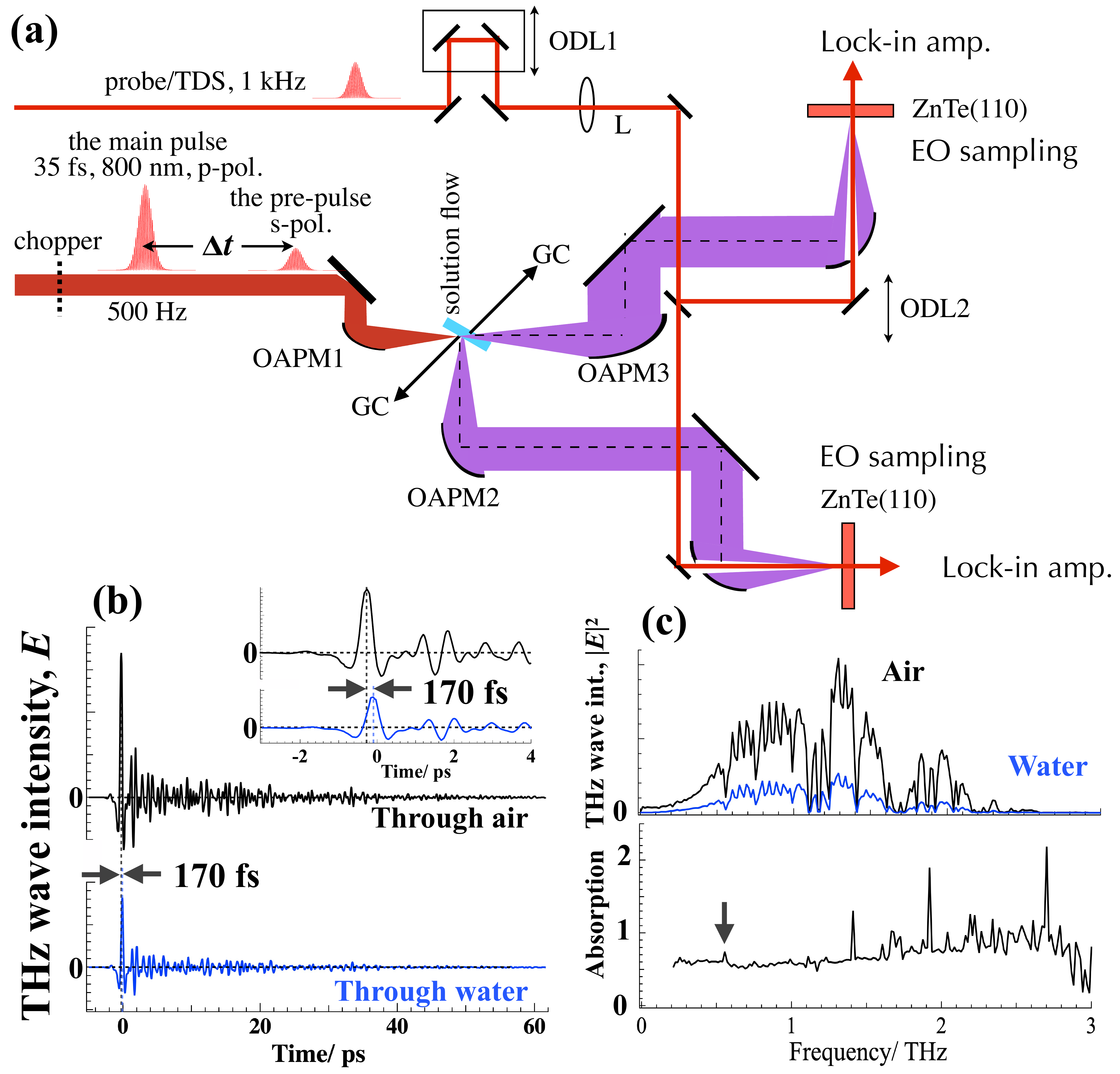}
\caption{(a) The experimental setup for THz wave with X-ray detections in the transmission and the reflection directions with time-domain spectroscopy (TDS). ODL1, optical delay for TDS. ODL2, optical delay for time matching between the transmission and the reflection directions. L, plano-convex lens ($f=50$~cm). The thickness of ZnTe(110) crystal for TDS was 1~mm. The parent focal lengths for the off-axis parabolic mirrors (OAPMs) were $f$ = 50.8~mm (OAPM1, 1-inch diameter), 101.6~mm (OAPM2, 2-inch diameter), 152.4~mm (OAPM3, 2-inch diameter), and two 101.6~mm (2-inch diameter with a hole in its centre for the probe) respectively. GC, Geiger counter. The inset shows the geometrical configuration of the flow, the laser incidence, and ablation phenomena in the nanosecond time delay. (b) TDS signals from a 1-mm thick ZnTe (110) crystal though air and the water flow at 60\textdegree incident angle. The inset represents the TDS signals near the zero delay. (c) Emission spectra in the THz frequency calculated from the TDS signals and calculated absorption spectrum of the water flow. The arrow indicates the absorption peak at 0.56 THz \cite{water1989}.}
\label{fig:exp_setup}
\end{figure}

Figure~\ref{fig:exp_setup}(a) shows the experiemtnal setup. The water flow nozzle system with two pipette tips and a reservoir, which has been introduced previously~\cite{hatanaka2016}, was used for creating a thin water flow. The flow thickness was measured at 17~$\pm$ 2~$\mu$m with a laser displacement meter (0.25~$\mu$m resolution, SI-T80, Keyence). The width of the flow is 8 mm at most, which is much wider than the laser focus size. More than 120 mL of water was being circulated by a pump (PMD-211, SANSO) with the flow rate at 70 mL/min during experiments. The flow angle and its position were controlled mechanically. The water flow position was monitored by a laser-based displacement sensor (0.2~$\mu$m resolution, LK-G80, Keyence).

%________________________________________fig 1

A pulsed femtosecond laser ($>$35 fs/transform-limited, 800 nm, Mantis, Legend Elite HE USP, Coherent, Inc.) was used. The laser beam was split into the pre-pulse (vertically-/ s- pol., 0.1 mJ/pulse) and the main pulse (horizontally-/ p- pol.). The remaining main pulse was split into the probe for THz time-domain spectroscopy (THz-TDS) and the main pulse (0.4 mJ/pulse) with a series of half-wave plates and polarization beam splitters (65-906, 47-048, Edmund Optics). The time delay between the pre-pulse and the main pulse was determined to be from 0 ns up to 15 ns using optical delays, and these two pulses were combined co-linearly for the double-pulse excitation using a cubic beam splitter (47-048, Edmund Optics). This combined beam passed through an optical chopper (3502, New Focus) which reduced the repetition rate from 1 kHz to 0.5 kHz for the lock-in measurements. The combined beams were tightly focused with a 1-inch off-axis parabolic mirror (OAPM, effective focal length $f$ = 50.8 mm, 47-097, Edmund Optics) onto the thin water film with the incident angle of 60\textdegree~to the flow normal.

\subsection{THz wave time-domain spectroscopy in the transmission/reflection directions}

THz wave emission in the reflection and the transmission directions, which are the front side and the back side of the flow, respectively, was recorded simultaneously by EO sampling with a $\langle 110\rangle$-oriented ZnTe crystal (1-mm thick, Nippon Mining~$\&$ Metals Co., Ltd.). This setup is an update from the previous one \cite{Hatanaka2018THz}. Briefly, the emission was collected by an OAPM ($f$ = 101.6~mm, MPD249-M01, ThorLabs) in reflection and by another OAPM ($f$ =152.4~mm, MPD269-M01, ThorLabs) in transmission. %The collected THz wave was then focused by another 2-inch OAPM (RFL=101.6~mm, MPD249H-M01, ThorLabs) with a hole, that allows the probe beam to pass through and co-propagates onto a ZnTe(110) crystal. 
The probe for TDS was split into two beams spatially with a d-shaped mirror (PFD05-03-M01, ThorLabs) which was set on a manual stage (ODL2) in order to match the time-zero between the reflection and the transmission directions.  The difference of the detection efficiencies in the reflection and the transmission directions was calibrated as follows; THz emission from a 1-mm thick $\langle 110\rangle$-oriented ZnTe crystal set before the OAPM1 was split into the transmission and the reflection components by an undoped silicon thin film as a THz beam splitter or bent by a conventional gold mirror placed at the water flow position. By measuring the TDS signal of THz wave in each direction, the difference of the detection efficiencies in the two directions was calibrated. With this system setup using the 1-mm thick $\langle 110\rangle$-oriented ZnTe crystals as detectors for EO sampling, the detectable frequency range extends to $\sim$ 3 THz at the highest~\cite{Wilke07}. 

%Polarization measurements in THz  emission spectroscopy were also carried out with wire-grids (MWG40FA-III, Origin, Ltd.). 

%The polarized probe beam propagated through a $\lambda$/4 plate and a Wollaston prism then was detected with a balanced photodiode (10$^5$ V/A, Model 2307, New Focus) and a lock-in amplifier (SR830, Stanford Research System).  which allowed not only a simultaneous detection of X-ray and THz wave but also THz in two different directions. 

\subsection{X-ray detection}

Geiger counters (SS315, Southern Scientific) with calibrated sensitivity were used for X-ray measurements at the front and the rear sides as described in detail elsewhere~\cite{Masim2016, Hsu2017}. The Geiger counters were set at $15^\circ$ to the flow normal and were placed 12~cm away from the water flow during the experiments. The vertically-polarized pre-pulse intensity in this experiemnt is low at 0.1~mJ/pulse and its irradiation does not induce detectable X-ray emission. All the experiments in this paper were carried out in air under atmospheric pressure (1 atm) at room temperature (296~K). Therefore, it is certain that the Geiger counter detects only X-ray, not $\alpha$- or $\beta$-rays.

%An undoped silicon thin film was used as the first step of THz-TDS for this newly developed system. It is a well-studied sample in THz region, suitable for testing the capabilities of the simhttps://www.overleaf.com/project/5ddc7eba82ae9e0001c38c12ultaneous detection and setting standards. A 1-mm~thick ZnTe (110) crystal was placed right before the OAPM1 as a THz wave emitter and a \textcolor{red}{100 $\mu$m-thick} undoped silicon was used as a THz wave splitter at the focal point to match the time zero between the reflection and the transmission directions. The thickness of the silicon thin film is confirmed experimentally by considering the peak time shift, as shown in Fig.~\ref{fig:exp_setup}b, since \textcolor{red}{THz peak shifts from propagating through materials with different refractive indices in TDS}\cite{dong17thickness}. On the basis of Snell's law, one can estimate the velocity of 1 THz traveling in the silicon by applying $\nu_{Si}$ = c/n$_{Si}$ = 8.8 $\times$ 10$^{7}$ m/s when n$_{Si}$ = 3.4 \cite{Dai04silicon}. With the observation of THz peak time shift, $\Delta$$t$_{Si}$ = 800 $\pm$ 0.6 fs, and the velocity of THz wave traveling through air the thickness of the silicon sample can be estimated to be d$_{Si}$ = $\nu_{Si}$ $\times$ $\Delta$$t$_{Si}$ = 71 $\pm$ 0.5 $\mu$m.

%Similar to the silicon film, 

%_________________________________fig 2
\begin{figure}[tb]
\centering\includegraphics[width=9.5cm]{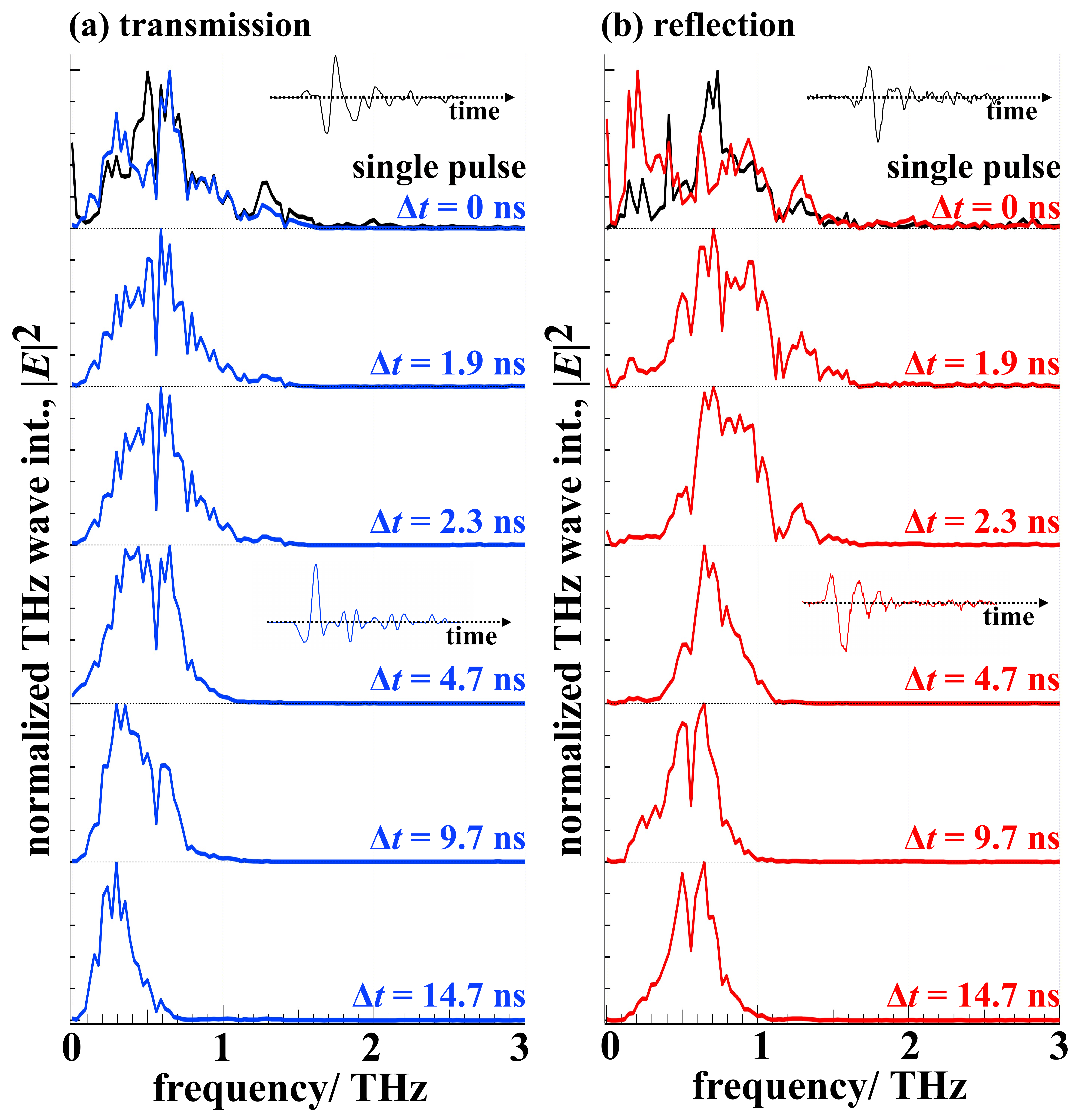}
\caption{Normalized THz emission spectra in (a) the transmission and (b) the reflection sides under the single pulse (in black on the top) or the double pulse excitation conditions.The vertical axes represent THz wave |\textbf{\textit{E}}|$^2$ intensity. The insets in the single pulse excitation case and $\Delta t$ = 4.7 ns represent the original signal wave-forms from -3 ps to 7 ps in TDS measurements.}
\label{fig:fig2}
\end{figure}
%_________________________________fig 3
\begin{figure}[tb]
\centering\includegraphics[width=11cm]{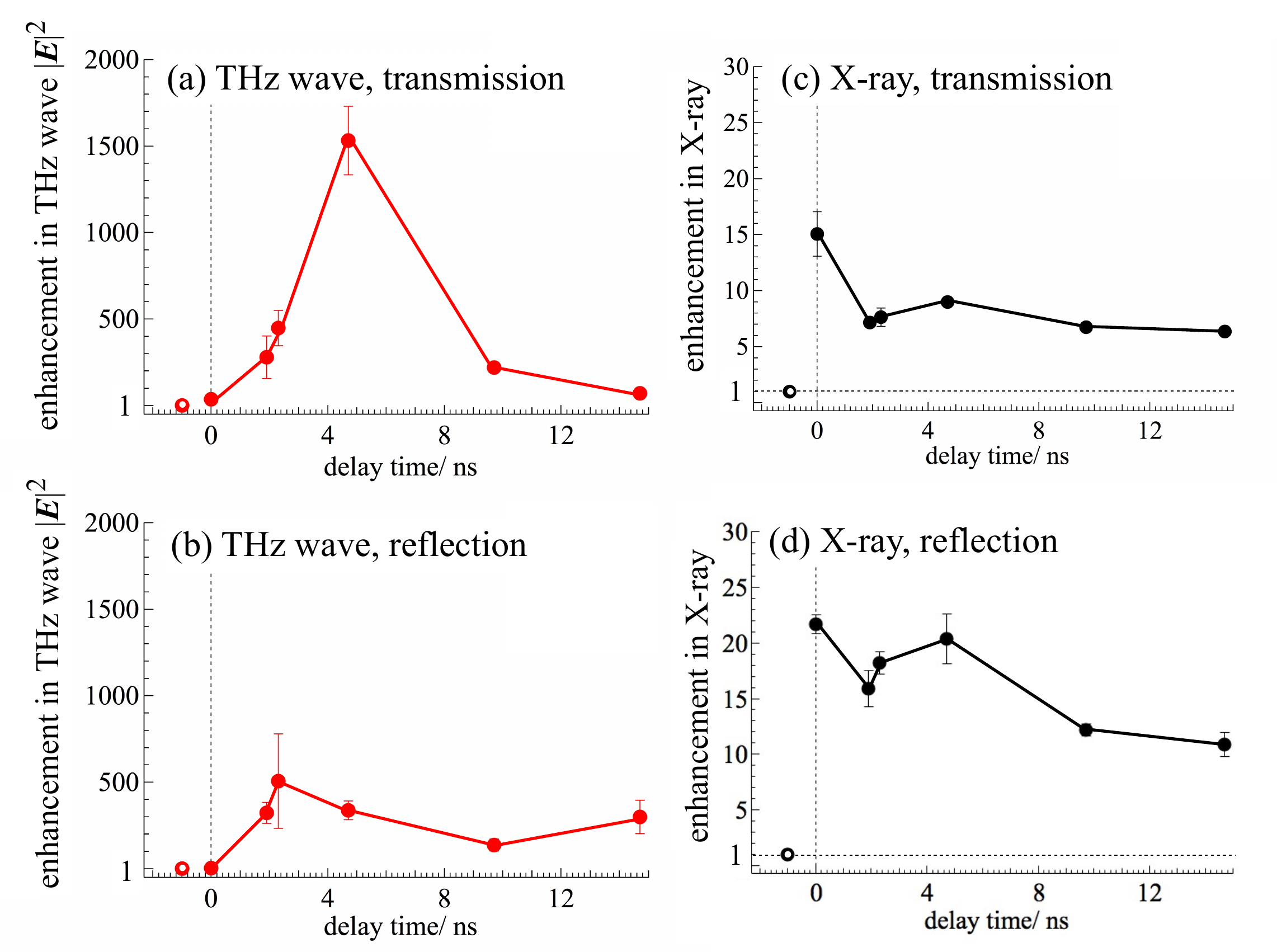}
\caption{Enhancements in THz wave emission in |\textbf{\textit{E}}|$^2$ (a, b) and X-ray (c, d) in the transmission and the reflection sides as a function of the delay time between the pre-pulse and the main pulse irradiations. The vertical axes are normalized with the emission intensity under the single pulse excitation condition (open circles).}
\label{fig:fig3}
\end{figure}

\section{Results}
\subsection{THz spectroscopy}

First, TDS measurements of THz wave emission from a 1-mm thick <110>-oriented ZnTe crystal set before the OAPM1 through the water flow were carried out. The residual laser pulse after the crystal was blocked by a black poly-propylene sheet placed right after the crystal.
Figure ~\ref{fig:exp_setup}(b) shows the TDS wave-forms through the water flow and air (without the flow). Here, the TDS signals were obtained from -5 ps to 61.6 ps every 0.03 ps.
The water flow displays a time shift, $\Delta$$t$$_{w}$ = 170 $\pm$ 0.6 fs, as compared with air.
%The water flow displays a time shift, $\Delta$$t$$_{w}$ = 170 $\pm$ 0.6 fs, as compared with air in TDS (Fig.~\ref{fig:exp_setup}(b)).
%, which is for the estimation of the flow thickness. 
With the refractive index of water at 1~THz, $n_{water}^{THz} = 2.12$\cite{Zhou19} and the light velocity in water, $v_{water} = c/n = 1.4\times 10^{8}$~m/s, considering the incident angle of 60\textdegree~to the flow normal, the thickness of the water flow is estimated to be $d_{w} = 14\pm 0.1~\mu$m. The water absorption spectrum shown in Fig.~\ref{fig:exp_setup}(c) is calculated by discrete Fourier-transform of the TDS signals. With the absorption intensity at 1 THz and the known absorption coefficient~\cite{Zhou19}, the flow thickness is estimated to be 15.7 $\mu$m, which is well matched with the estimated values. In the spectrum, there are a few sharp lines (e.g., indicated by an arrow at 0.56 THz), which can be assigned to water monomer rotational energy levels~\cite{Xin06vapor}. The flow thickness estimated is much smaller than the wavelength in the THz wave emission ($\sim$300 $\mu$m), therefore interference due to internal multiple reflection in the flow as considered in the previous report~\cite{ZhangarXiv19} is negligible.

%\textcolor{blue}{The spectrscopically well-resolved absorption peak of water at 0.56~THz (Fig.~\ref{fig:exp_setup}(c)) shows possibility to measure absorbance from small $\sim\lambda/30$ sub-wavelength water flow by THz radiation produced by fs-laser pulses.}

Figure 2 shows normalized spectra of THz emission in (a) transmission and (b) reflection from the water flow irradiated by the laser pluses.
Here, the experimental condition is slightly different from that in Fig.,1; the TDS signals were obtained from -10 ps to 24 ps every 0.07 ps.
%Figure 2 shows normalized THz emission spectra in (a) transmission and (b) reflection. 
Under the single pulse excitation condition, the spectra are similar to each other in both of the directions. Under the double pulse excitation condition, the series of the spectra commonly show their dynamic peak shifts toward the lower frequency in the nanosecond time range (to the maximum delay at 14.7 ns). In this time range, after the decay of the plasma by the pre-pulse irradiation, laser ablation phenomena such as transient surface roughness with ripple formation, shock-wave expansion, and micro-droplet/mist ejection in the direction parallel to the flow normal as shown in the inset of Fig.1(a) {\cite{hatanaka2002}} are already initiated. This is reconfirmed in the observation of the water absorption peak at 0.56 THz (the arrow in Fig.1(c)) even for the reflection direction (Fig.~2(b)). 
%\textcolor{red}{

%The insets in $\Delta t = 4.7$~ns represent the original signal wave-forms in TDS with the vertical axis of THz wave $\textbf{\textit{E}}$ intensity as a function of time, which shows that the phases in TDS are not the same in the transmission and the reflection directions in the nanoseconds time delay.
%}

\subsection{THz enhancement}

Figure 3 shows dynamic changes of the enhancement factors of THz wave emission in |\textbf{\textit{E}}|$^2$ over the frequency range in 0-3~THz (a, b) and X-ray emission (c, d) normalised to the emission intensity under the single pulse irradiation condition in transmission and reflection, respectively. In the previous report \cite{Hatanaka2018THz}, we estimated such enhancement factor only based on the peak intensity of the original TDS waveform as $\textbf{\textit{E}}$-filed intensity. However, the estimation in the current manner is more reliable. With the pre-pulse (s-pol.) and the main pulse (p-pol.) intensities at 0.1~mJ/pulse and 0.4~mJ/pulse, respectively, and the transform-limited pulse at 35~fs, the enhancement of the THz emission shows its highest at the 2-5 ns window, while that of the X-ray shows its peak at 0~ns. Evidently, this data set shows the different enhancement mechanisms for both of the emission.

In the case of X-ray, the pre-pulse irradiation induces the plasma formation and its fast decay at around 0~ns delay time, then the main pulse irradiates this pre-excited region leading to jump in plasma temperature and electron density. High temperature is the cause of X-ray emission and its enhancement.In the case of THz emission which may depend on temperature gradient,such hot plasma formation may interfere THz emission since the critical density of plasma~\cite{Chen1984}, $n_c = \omega_p^2m_e\varepsilon_0/e^2$, where $n_c$, $\omega_p$,  $\varepsilon_0$,  $m_e$ and $e$ are the critical electron density, plasma frequency, permittivity, electron mass and charge, respectively, makes the excited region opaque. %is too high for THz wave at lower frequency range, then its emission is hindered. 
In the nanoseconds time delay after the pre-pulse irradiation, various laser ablation phenomena described above have already started. The acoustic space-time scaling of sub-$\mu$m features formed in a few tens of $\mu$s as described above. Importantly, the highest enhancement in THz wave emission under the double pulse excitation is $1.5\times 10^3$ times in intensity |\textbf{\textit{E}}|$^2$ at 4.7 ns delay in the transmission direction, while the X-ray enhancement is only 15-22 times at the highest. 
\begin{figure}[tb]
\centering\includegraphics[width=8.5cm]{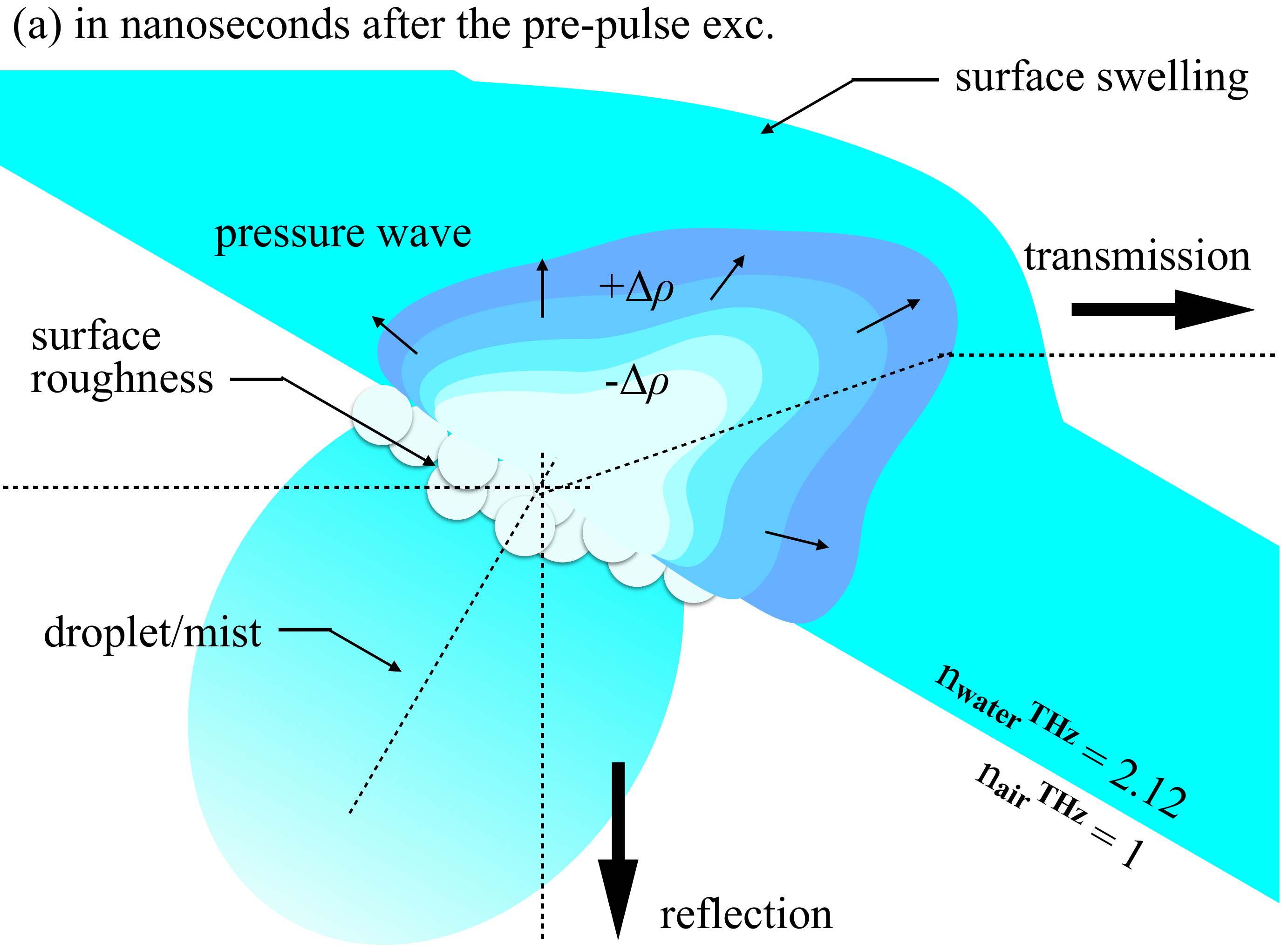}
\centering\includegraphics[width=8.5cm]{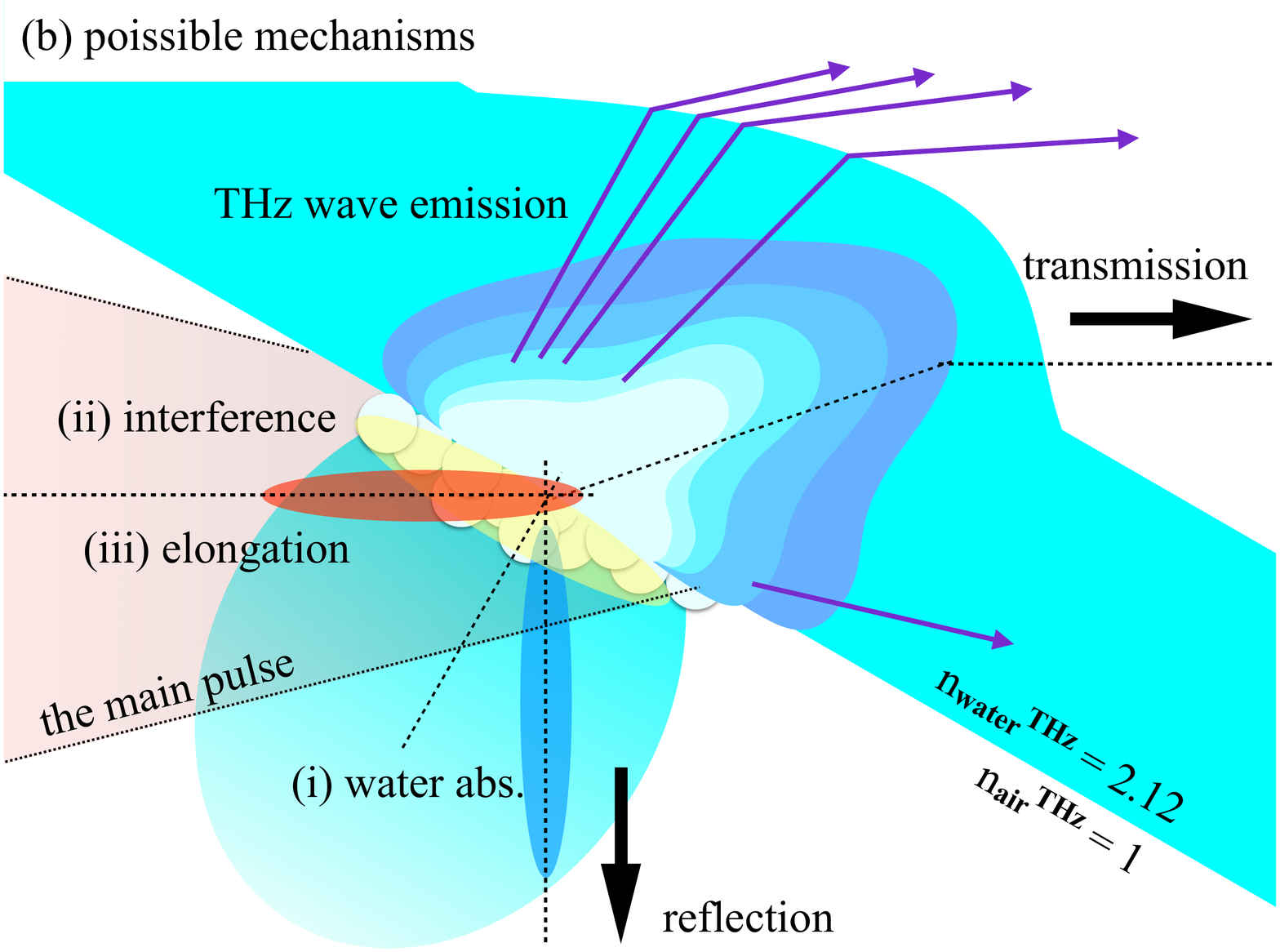}
\caption{(a) A schematic diagram for the water flow structure (top view) at the delay time in nanoseconds after the pre-pulse irradiation. (b) Possible mechanisms for the interaction of the main pulse with the structured water flow and resultant THz wave emission.}
\label{f-THz}
\end{figure}

%\begin{figure}[tb]
%\centering\includegraphics[width=14cm]{f-THz.pdf}
%\caption{(a) Schematics of the pre-pulse irradiation to the water flow. It leads to surface nano- to micro-scale roughness formation in the nanosecond time range. The angle of the laser incidence $\theta_i = 60^\circ$ is chosen for the maximum X-ray generation. (b) Schematics of formation  of modified region throughout the water flow at the arrival of the main-pulse after $\Delta t > 4$~ns time delay. (c) THz frequency for the reflection as a function of the critical plasma density. (d) Phase change for the attenuated total reflection (ATR)  at the water ($n_i=2.12$ in THz frequency)- air ($n_t =1$) interface; the phase difference is also plotted for $n_i=1$-to-$n_t=0.9$ (inside a rarefied channel (b)). The polarising Brewster $\theta_B = \arctan(n_i)$ and the critical $\theta_c = \arcsin(n_i)$ angles are marked.     }
%\label{f-THz}
%\end{figure}

\section{Discussion}
\subsection{Effects of the pre-pulse irradiation}

By considering the ablation phenomena at nanoseconds time delays after the pre-pulse, the observed dynamic spectral shift toward the low frequency, the phase inversion, and the giant enhancement of THz emission 
%\sout{\purple{and its ellipticity}} 
are discussed from the viewpoints of THz wave absorption by water, surface structural modifications of the flow such as roughness/ripple formation, mist ejection, and swelling, an elongation of THz wave source, THz wave emission characteristics, 
%attenuated total reflectance (ATR) due to refractive index transients, 
and resonance absorption. 

Based on the knowledge on laser ablation of liquids under the similar experimental conditions \cite{hatanaka2002}, the structural modifications of the water flow induced by the pre-pulse irradiation, nano- to micro-scale roughness/ripples, droplet/mist ejection, and pressure wave propagation with the swelling on the rear surface are considered as shown in Fig.4(a). The scale length is governed by the speed of sound $v_s=1.481$~km/s (under normal conditions) and the delay time after the pre-pulse irradiation, $\Delta t$. At the delay time of 4.7 ns, for instance, such pressure wave travels $v_s\times\Delta t\approx 7~\mu$m , which is comparable with the thickness of the water flow. Hence, the pressure induces the increase of mass density $+\Delta\rho$ due to the compression and causes the rarefaction $-\Delta\rho$ at the central part which leads to cavitation on longer time scale.   

%in Fig.5(a)) after the formation of free carrier and of breakdown plasma alongside with the pressure-induced rareficaton (reduced mass density leading to cavitation). These processes contribute to the formation of regions with the density gradient ($-\Delta$$\rho$ and $+\Delta$$\rho$) and the rear surface swelling on the flow as shown in Fig.5(a). Under the geometrical conditions described, the features observed are quantitatively discussed as bellow. 

\subsection{Spectral peak shift}

One straightforward idea is that droplet/mist ejection in the detection path for the THz wave emission in the transmission and the reflection directions may cause additional absorption for THz wave emission as shown in Fig.4 (b). The water absorption spectrum shows its higher intensity at the higher frequency as reported previously~\cite{Allen06}, which will cause an apparent peak shift toward the lower frequency as the droplet/mist ejection takes place.

Second, interferometric effects for THz wave emission due to the transient surface roughness/ripple formation on the flow surface are considered as shown in Fig.4 (b). It was reported that laser excitation of LiNbO$_3$ by transient grating controls THz emission spectra within the range of 0.5-3 THz~\cite{Stepanov04}. Similar studies with InAs proved that periodic structures of THz wave emission source show the peak shift toward the lower frequency as the period increases from 190~$\mu$m to 360 $\mu$m ~\cite{Hiratsuka07}. Under the current experimental condition for the pre-pulse irradiation to induce laser ablation, transient surface roughness is considered to be composed of different orders of spatial frequencies from a few-tens of nanometers to micrometers~\cite{hatanaka2002}. Such surface structures can cause the dynamic spectral shift  under the delayed main pulse irradiation.

\subsection{Giant enhancement in THz emission}

As shown in Fig.~3(c,d), X-ray intensity, not only THz wave emission, also increases to some extent at the same delay time of 4.7 ns. These enhancements in the both emissions in THz wave and X-ray indicate that the absorption of the main pulse increases effectively and that the density and the temperature of electrons increases in the self-generating plasma by the main pulse irradiation. The pre-pulse irradiation induces the droplet/mist formation and its expansion outward with time. Then, the delayed main pulse induces the self-generating plasma with a larger volume in the expanding droplet/mist, which results in the enhanced absorption. Or, due to the flow surface with the structural modulations such as nano- to micro-scale roughness/ripples induced by the pre-pulse irradiation, the effective absorption efficiency of the main pulse increases through multiple scatterings. Furthermore, owing to such effective absorption, the electron density in the self-generating plasma increases to its critical density for the wavelength of the main pulse and the resonance absorption effect may become a dominant process for the enhanced emission of THz wave and X-ray.

%Considering the giant increase of the THz pulse energy compared with that of X-ray intensity, additional mechanisms for the increase of the THz pulse energy should exist.
%The phase change of such waveguiding in the ATR central channel was qualitatively modeled with $n_i\approx 1.1$ and $n_t\approx 0.9$ (Fig.~\ref{f-THz}(d)). Phase change at angle of incidence $\theta_B<\theta_i<\theta_c$ is $\pi$ (0.5 waves; $\theta_B$ is the polarising Brewster angle) and is $0^\circ$ for $\theta_i < \theta_B$.

%\textcolor{brown}{One possible explanation of the observed giant enhancement of THz wave emission (Fig.~\ref{fig:fig3}) is the formation of the rarefied central channel (less water, $-\Delta\rho$ in Fig.5(a)) prepared by the pre-pulse irradiation. By the main-pulse irradiation, the plasma side-walls are formed and surrounds the channel as shown in Fig.~5(b). The walls become opaque (metallic) and reflect the main pulse generating THz wave toward the optical axis into the transmission channel. This qualitative scenario of THz wave enhancement due to better channeling of THz wave radiation into the transmission channel is consistent with  experimental data. }

%\vspace{5mm}
In summary, we tentatively consider that the giant THz wave enhancement at 4.7 ns is attributed to the self-generating plasma with the larger volume in the expanding droplet/mist induced by the pre-pulse irradiation.
In addition, the modification of the rear surface induced by the pre-pulse irradiation as shown in Fig. \ref{f-THz} may contribute to the enhancement observed in the transmission side at the delay time of 4.7 ns.
% In addition, the reduction of ATR effects at the swelled rear surface of the flow may contribute to the enhancement observed in the transmission at the delay time of 4.7 ns. 
%\purple{\sout{The polarization status can be determined by the transmission of the THz wave emission through the rear surface.
%The change in transmission of THz wave through the modified surface could be the origin of the elliptic polarization observed at the delay time of 4.7 ns.} }
To reach more quantitative descriptions of the light-matter interaction for the described experiment, a fast optical imaging will be added to reveal the geometry and its changes during the laser ablation of the water flow by the pre-pulse irradiation. Timing of opening of a channel with density gradient in the film and extent of nano-to micro-droplets/mist and swelling are necessary to quantify and further validate separate contributions of discussed mechanisms to the enhancement of THz wave emission.

\section{Conclusions and Outlook}

This paper describes the simultaneous emission of THz wave and hard X-ray from distilled water when it is irradiated by focused double pulse of femtosecond laser. THz wave emission spectra show their dynamic peak shifts toward the lower frequency with the polarization change to elliptical, which has been discussed in the viewpoints of laser ablation induced by the pre-pulse irradiation. THz wave emission intensity |\textbf{\textit{E}}|$^2$ increases more than $1.5\times 10^3$ times   when it is compared to the emission intensity under the single pulse excitation condition, which is recalculated to be 10-times higher than the THz wave emission from the conventional $\langle 110\rangle$-oriented ZnTe crystal. Further enhancements are surely expected with chirped pulse excitation \cite{hatanaka2008, Savelev18} or with the addition of solutes \cite{HatanakaChem299, hatanakaapl2008}. Optimisation of the double pulse irradiation and target solutions will be focus of next studies for advancements of THz sources.

\small\section*{Funding}

TN is grateful for the support by JSPS KAKENHI Grant Number 17K05086. TY is grateful for the support by Grant-in-Aid for Scientific Research for Fostering Joint International Research (B) (18KK0159), SJ is grateful for the support via the Australian Research Council Discovery project DP190103284.
KH is grateful for the supports by the Ministry of Science and Technology (MOST) of Taiwan (107-2112-M-001-014-MY3), the Cooperative Research Program of "Network Joint Research Center for Materials and Devices",  Nanotechnology Platform, and the Collaborative Research Projects of Laboratory for Materials and Structures, Institute of Innovative Research, Tokyo Institute of Technology.

\small\section*{Disclosures}

The authors declare that there are no conflicts of interest related to this article.

%%%%%%%%%%%%%%%%%%%%%%% References %%%%%%%%%%%%%%%%%%%%%%%%%

%Add references with BibTeX or manually.
%\cite{Zhang:14,OSA,FORSTER2007,Dean2006}

%%%%%%%%%% If using BibTeX:
\bibliography{THz}

%%%%%%%%%% If preparing manually:
% \newcommand{\enquote}[1]{``#1''}

% \bibitem{Zhang:14}
% Y.~Zhang, S.~Qiao, L.~Sun, Q.~W. Shi, W.~Huang, L.~Li, and Z.~Yang,
%   \enquote{Photoinduced active terahertz metamaterials with nanostructured
%   vanadium dioxide film deposited by sol-gel method,}
%   {\protect\JournalTitle{Optics Express}} \textbf{22}, 11070--11078 (2014).

% {Optical Society}, \enquote{{OSA Publishing},}
%   \url{http://www.osapublishing.org}.

% \bibitem{FORSTER2007}
% P.~Forster, V.~Ramaswamy, P.~Artaxo, T.~Bernsten, R.~Betts, D.~Fahey,
%   J.~Haywood, J.~Lean, D.~Lowe, G.~Myhre, J.~Nganga, R.~Prinn, G.~Raga,
%   M.~Schulz, and R.~V. Dorland, \enquote{Changes in atmospheric consituents and
%   in radiative forcing,} in \enquote{Climate Change 2007: The Physical Science
%   Basis. Contribution of Working Group 1 to the Fourth assesment report of
%   Intergovernmental Panel on Climate Change,}  S.~Solomon, D.~Qin, M.~Manning,
%   Z.~Chen, M.~Marquis, K.~B. Averyt, M.~Tignor, and H.~L. Miler, eds.
%   (Cambridge University Press, 2007).

% \end{thebibliography}

\end{document}